\documentclass[a4paper,superscriptaddress,floatfix,showpacs,aps]{revtex4}
\usepackage[english]{babel}
\usepackage[utf8]{inputenc}
\usepackage{amsmath}
\usepackage{graphicx,epstopdf}
\usepackage{amssymb,epsfig,color,textcase}
\usepackage{hyperref}

\begin{document}

\title{Mechanism of ferromagnetic ordering of the Mn chains in CaMnGe$_2$O$_6$ clinopyroxene}

\author{F. V. Temnikov}
\affiliation{Ural Federal University, Mira Str. 19, 620002 Ekaterinburg, Russia}

\author{E. V. Komleva}
\affiliation{Institute of Metal Physics, Russian Academy of Science, S. Kovalevskaya Str. 18, 620041 Ekaterinburg, Russia}
\email{komleva@imp.uran.ru}

\author{Z. V. Pchelkina}
\email{pchelkzl@mail.ru}
\affiliation{Institute of Metal Physics, Russian Academy of Science, S. Kovalevskaya Str. 18, 620041 Ekaterinburg, Russia}
\affiliation{Ural Federal University, Mira Str. 19, 620002 Ekaterinburg, Russia}

\author{S. V. Streltsov}
\email{pchelkzl@mail.ru}
\affiliation{Institute of Metal Physics, Russian Academy of Science, S. Kovalevskaya Str. 18, 620041 Ekaterinburg, Russia}
\affiliation{Ural Federal University, Mira Str. 19, 620002 Ekaterinburg, Russia}

\date{\today}

\begin{abstract}
The electronic and magnetic properties of clinopyroxene CaMnGe$_2$O$_6$ were studied using density function calculations within the GGA+U approximation. It is shown that anomalous ferromagnetic ordering of neighboring chains is due to a ``common-enemy'' mechanism. Two antiferromagnetic exchange couplings between nearest neighbours within the Mn-Mn chain and interchain coupling via two GeO$_4$ tetrahedra suppress antiferromagnetic exchange via single GeO$_4$ tetrahedron and stabilize ferromagnetic ordering of Mn chains.
\end{abstract}

\maketitle

Pyroxenes are a large group of rock-forming minerals widespread in the Earth's crust and its upper mantle~\cite{Anderson-book}. These materials are important not only for the geoscience, but also rather interesting for the condensed matter physics, especially those of them, which contain transition metal ions. For instance, some of such pyroxenes show orbitally assisted Peierls effect and opening of the spin gap~\cite{Konstantinovic-2004,Wezel-2006,Streltsov-2008,Feiguin-2019}, others demonstrate cooperative Jahn-Teller distortions~\cite{Ohashi-1987} or rare combination of ferromagnetism and insulating behaviour~\cite{Vasiliev-2005}, and there are even multiferroics among pyroxenes~\cite{Jodlauk-2007}.  

Discovery of magneto-electric effect in pyroxenes with general formula ATM(Si,Ge)$_2$O$_6$ (where TM is a transition metal ion and A can be alkali or alkaline earth metals) resulted in intensive studies of their magnetic structure and its coupling with electronic properties and lattice distortions. Depending on particular choice of TM or A ions there were observed very different types of magnetic structures in pyroxenes including collinear antiferromagnets, commensurate and incommensurate spin spirals and even ferromagnets~\cite{Vasiliev-2005,Nenert-2009,Redhammer-2008,Redhammer-2011}. Such a variety of magnetic orderings is due to low dimensionality of the crystal structure and frustration effects intrinsic for the pyroxene lattice. 

\begin{figure}
\includegraphics[width=0.37\textwidth]{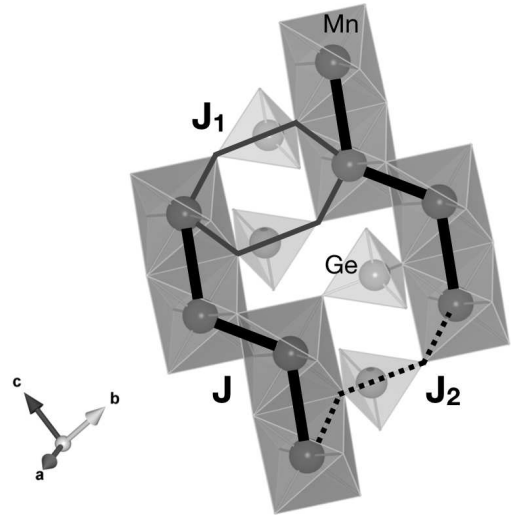}
  \caption{The MnO$_6$ zigzag chains connected through GeO$_4$ tetrahedra in CaMnGe$_2$O$_6$.  The bold solid, thin solid and dotted lines correspond to exchange paths $J$, $J_1$, and $J_2$ as suggested in~\cite{Ding-2016}}
  \label{fig1}
\end{figure}

In pyroxenes transition metal ions are in the ligand octahedra, which form one dimensional (zigzag) chains sharing their edges, see Fig.~\ref{fig1}. The strongest exchange coupling is typically within these chains.  The chains are connected by (Si/Ge)O$_4$ tetrahedra and this provides various interchain couplings, which could make a whole spin system frustrated. The density functional theory (DFT) was shown to be useful in analysis of the exchange interaction and helped to explain magnetic properties for a number of pyroxenes~\cite{Streltsov-2008,Janson-2014,Lee-2014,Ding-2018}.

In the present paper we perform DFT calculations to study electronic and magnetic properties of CaMnGe$_2$O$_6$, which magnetic structure was recently refined~\cite{Ding-2016}. It was found that it can be described as antiferromagnetic chains running along $c$ direction, ordered, however, ferromagnetically. It was known from long ago that any ferromagnetic coupling is rather untypical for insulating strongly correlated materials, since it is due to overlap between half-filled and empty orbitals, which scales as $1/U^2$ with Hubbard $U$, while conventional exchange interaction between half-filled orbitals behave as $1/U$~\cite{Goodenough,Streltsov-2017}. Thus, it is important to find out the mechanism resulting in ferromagnetic ordering in CaMnGe$_2$O$_6$.


DFT calculations were performed in the Vienna ab initio simulation package (VASP)~\cite{vasp2} with exchange-correlation potential chosen as proposed in~\cite{exchcorrp2}. Strong Coulomb correlations were taken into account within the GGA+U approximation~\cite{LSDA}. On-site Hubbard interaction $U$=4.5~eV was taken along with Hund's coupling parameter $J_H$=0.9~eV~\cite{Streltsov-2008,Streltsov2014d}. The plane-wave energy cutoff was chosen to be 520 eV. Fine $6\times6\times12$ \emph{k-}mesh was used in the calculations and \emph{k}-space integration was performed by tetrahedron method. The convergence criterion for the total energy was chosen to be 10$^{-7}$ eV.

\begin{figure}[t]
  \begin{center}
  \includegraphics[clip=false,angle=270, width=0.5\textwidth]{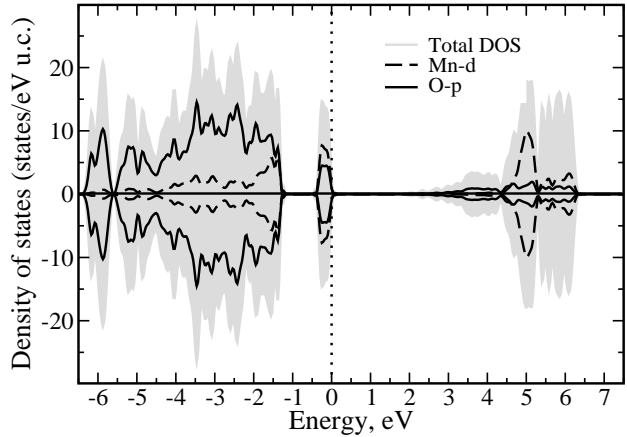}\\
  \caption{Total and partial (Mn and O) densities of states calculated in AFM1 ground state of CaMnGe$_2$O$_6$ within GGA+U approximation. The Fermi level corresponds to zero energy}
  \label{DOS}
  \end{center}
\end{figure}

We start by analyzing electronic structure of CaMnGe$_2$O$_6$, as obtained in the GGA+U calculations for the experimental ground state magnetic structure (as it will be explained below the same magnetic structure, AFM1, corresponds to the total energy minimum in the GGA+U calculations). One can see from the density of states plot, presented in Fig. \ref{DOS}, that CaMnGe$_2$O$_6$ is an insulator with the band gap of $\sim$ 2~eV. Narrow band of $\sim$ 0.5 eV width right below the Fermi level is mostly formed by the Mn-$3d$ and O-$p$ states. Valence band has an additional gap of $\sim$ 1 eV. 

\begin{figure}[t]
  \begin{center}
  \includegraphics[clip=false,width=0.5\textwidth]{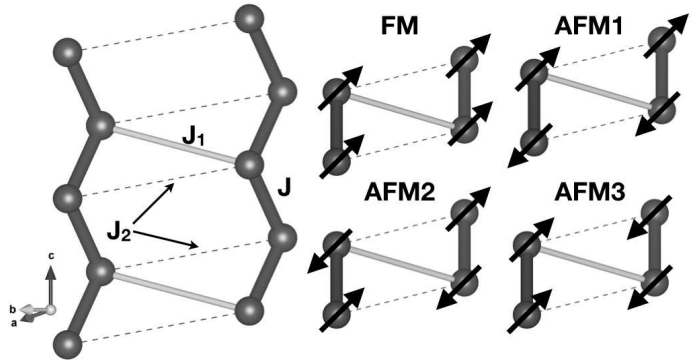}\\
  \caption{Four spin configurations used in the total energy calculations. The thick black solid line corresponds to the exchange along the chain $J$, the others describe interchain interactions - thin grey solid line shows $J_1$ path (via two GeO$_4$ tetrahedra), dashed line -- $J_2$ (via one GeO$_4$ tetrahedron)}
  \label{fig3}
  \end{center}
\end{figure}

In order to find origin of ferromagnetic ordering of Mn chains we calculated exchange parameters $J_{ij}$ of the Heisenberg model, which was written in the following form:
\begin{equation}
H=\sum_{ij}{J_{ij} \mathbf{S}_i \mathbf{S}_j},
\label{eq1}
\end{equation}
where $i$ and $j$ numerate lattice sites. The total energy method as realized in the JaSS code~\cite{Jass} was applied to calculate intrachain, $J$, and interchain exchange parameters $J_1$ (via two GeO$_4$ tetrahedra) and $J_2$ (via one GeO$_4$ tetrahedron). Four different magnetic configurations presented in Fig.~\ref{fig3} were used.

According to our calculations AFM1 configuration has the lowest total energy so it can be considered as the ground state. In this configuration neighboring spins in the chain are ordered antiferromagnetically while in neighboring chains - ferromagnetically. This agrees with experimental results~\cite{Ding-2016}. Calculated magnetic moment on Mn$^{2+}$ ions (electronic configuration 3$d^5$, $S=5/2$) for the ground state magnetic order was found to be 4.6 $\mu_B$ that is in line with experimentally obtained 4.41 $\mu_B$~\cite{Redhammer-2008} and 4.71 $\mu_B$~\cite{Ding-2016}.

It is interesting, that in spite of ferromagnetic order of neighboring chains all isotropic exchange parameters turned out to be antiferromagnetic. The dominating exchange parameter is intrachain exchange $J$=3.6 K, it is 3 times larger than $J_1$=1.2 K and 10 times larger than $J_2$=0.3 K. In general, both $J$ and $J_1$ determine magnetic structure of the investigated pyroxene and suppress weak AFM $J_2$ exchange making spins in neighboring chains to order ferromagnetically (see Fig.~\ref{fig3}). This is exactly what we see in AFM1 configuration and this is consistent with the experimental magnetic structure given in Ref.~\cite{Ding-2016}. It allows us to answer the question raised in the beginning: these are two strong antiferromagnetic exchange interactions, which drive ferromagnetic arrangement of Mn chains in CaMnGe$_2$O$_6$. Obtained values of isotropic exchange parameters also show that CaMnGe$_2$O$_6$ can be considered as a quasi-one-dimensional magnet. Indeed, though $J$, $J_1$, and $J_2$ exchange paths form triangle network, $J_1=1.2$ K is much larger than  $2J_2=0.6$~K and thus frustration is mostly suppressed in CaMnGe$_2$O$_6$. This is contrast to NaFeGe$_2$O$_6$, where 2$J_2$=4.2 K is close to $J_1$=3.8 K indicating strong frustrations~\cite{Ding-2018}.  
\begin{figure}[t]
  \begin{center}
  \includegraphics[clip=false,width=0.5\textwidth]{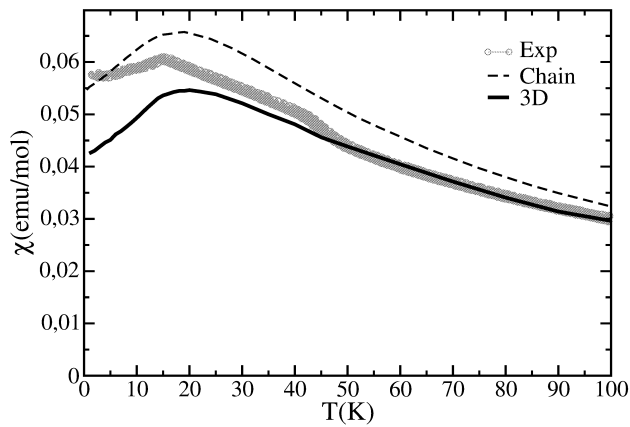}\\
  \caption{Fig. 4. Fit of the magnetic susceptibility of CaMnGe$_2$O$_6$ within 3D spin model including $J$, $J_1$, $J_2$ (solid line). The susceptibility of the spin chain with only $J$ exchange is shown by the dashed line. Experimental data are taken from Ref.~\cite{Ding-2016}}
  \label{fig5}
  \end{center}
\end{figure}

Using the mean-field theory we estimated Curie-Weiss temperature, 
\begin{equation}
\theta=-\frac{2S(S+1)}{3}\sum_{i}{z_iJ_i}=-\frac{35}{3}(J+J_1+2J_2),
\label{eq2}
\end{equation}
where $z_i$ is the number of exchange paths per Mn site. For exchange parameters obtained for $U$=4.5 eV expression (\ref{eq2}) results in $\theta_{calc}$=$62.8$ K, while experimental value  is $\theta_{exp}=35.1$ K~\cite{Ding-2016}. Taking into account that the mean field approach often overestimates $\theta$ by 2-3 times, one sees that calculated $\theta$ agrees with experimental estimation. 

We proceed further comparing temperature dependence of magnetic susceptibility obtained using calculated in the GGA+U approximation exchange parameters with experimental $\chi(T)$. For this  we used classical Monte Carlo simulations of (\ref{eq1}) using SPINMC algorithm of the ALPS package~\cite{alps}. The $L\times L\times L$ finite lattices with $L$ up to 8 and periodic boundary conditions were used. The magnetic susceptibility for the 3D spin model including three exchange paths $J$, $J_1$, and $J_2$ is shown in Fig.~\ref{fig5} (solid line). The susceptibility for the isolated spin chain with the main exchange $J$ (dotted line) is also shown Fig.~\ref{fig5}. The comparison of magnetic susceptibilities with experiment shows that the isolated chain model does not match experimental data for CaMnGe$_2$O$_6$ neither in absolute values nor in slope. Small kink at $\sim$45 K in experimental $\chi(T)$ is attributed to small ferrimagnetic Mn$_3$O$_4$ impurities~\cite{Ding-2016}.

Finally, in order to check stability of the results, we repeated the GGA+U calculations for slightly different choice of Hubbard $U$ parameter. All results are summarized in Tab.~1. They basically show the same tendency: the strongest is the exchange interaction along Mn chains, while the second largest exchange coupling is via two GeO$_4$ tetrahedra ($J_1$). 
\begin{table}[t]
\centering
\caption{Calculated exchange parameters $J$, $J_1$ and $J_2$ (in K) for various $U$. Hund's intra-atomic exchange $J_H$ was fixed at 0.9 eV.}
\begin{tabular}{|c|c|c|c|}
\hline
$J_i$& $U$=3.5 eV & $U$=4.5 eV& $U$=5.5 eV\\
\hline 
$J$   & 4.2 & 3.6 & 3.1\\  \hline
$J_1$ & 1.5 & 1.2& 0.9\\  \hline
$J_2$ & 0.4 & 0.3 & 0.3\\  \hline
\end{tabular}
  \label{tabl2}
\end{table}

To summarize, electronic structure and magnetic properties of CaMnGe$_2$O$_6$ were studied using the GGA+U calculations. The calculated values of exchange interaction parameters allow to explain the experimentally observed magnetic structure with antiferromagnetic interaction within the zigzag Mn chains and ferromagnetic ordering of these chains. The Monte Carlo simulation of magnetic susceptibility within 3D spin model with calculated exchange parameters agrees with experimental data much better than the one for isolated spin chains stressing importance of interchain coupling. The obtained values of exchange interactions also indicate that in CaMnGe$_2$O$_6$ magnetic frustration is weak. Such a weak frustration could explain the commensurate collinear antiferromagnetic structures common for Ca$^{2+}$-bearing pyroxenes. 

\begin{acknowledgements}
This work was supported by the Russian Science Foundation through RSF 17-12-01207 research grant.
\end{acknowledgements}

\end{document}